# A New Momentum-Integrated Muon Tomography Imaging Algorithm

JungHyun Bae,* Rose Montgomery*, Stylianos Chatzidakis[†]

*Oak Ridge National Laboratory, 5200, 1 Bethel Valley Rd, Oak Ridge, TN 37830, baejh@ornl.gov
[†]School of Nuclear Engineering, Purdue University, 363 North Grant Street, West Lafayette, IN 47907

## INTRODUCTION

For decades, the application of muon tomography to spent nuclear fuel (SNF) cask imaging has been theoretically evaluated and experimentally verified by many research groups around the world, including Los Alamos National Laboratory in the United States, Canadian Nuclear Laboratory in Canada, the National Institute for Nuclear Physics in Italy, and Toshiba in Japan [1–3]. Although monitoring of SNF using cosmic ray muons has attracted significant attention as a promising nontraditional nondestructive radiographic technique, the wide application of muon tomography is often limited because of the natural low cosmic ray muon flux at sea level: ~$10^2$ m$^{-2}$ min$^{-1}$sr$^{-1}$ [4]. Recent studies suggest measuring muon momentum in muon scattering tomography (MST) applications [5–7] to address this challenge. Some techniques have been discussed [8–11]; however, an imaging algorithm for momentum-coupled MST had not been developed. This paper presents a new imaging algorithm for MST which integrates muon scattering angle and momentum in a single M-value. To develop a relationship between muon momentum and scattering angle distribution, various material samples (Al, Fe, Pb, and U) were thoroughly investigated using a Monte Carlo particle transport code GEANT4 simulation. Reconstructed images of an SNF cask using the new algorithm are presented herein to demonstrate the benefit of measuring muon momentum in MST. In this analysis a missing fuel assembly (FA) was located in the dry storage cask.

## MUON SCATTERING TOMOGRAPHY USING POINT OF CLOSEST APPROACH (POCA)

When a muon interacts with matter, it is deflected, and its flight direction changes because of continuous Coulombic interactions with nuclei and electrons: this is known as *multiple Coulomb scattering* (MCS). Although improved models to reconstruct MCS trajectory in the scattering medium have been developed based on Bayesian estimation and maximum likelihood expectation minimization [12][13], PoCA is still widely used for muon tomography as an imaging algorithm because it is fast and efficient [14]. The PoCA algorithm finds a single scattering point for each muon trajectory history, and it is geometrically calculated from a midpoint of the shortest perpendicular line between incoming and outgoing muon trajectories. A PoCA voxel, $V_{n3}(l,m,n)$, is assigned to the associated 3D coordinate ($P_x$, $P_y$, $P_z$)

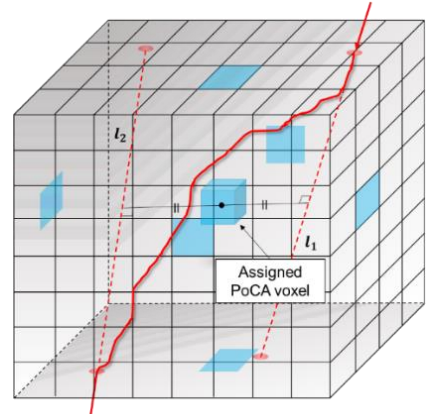

Fig. 1. Visualized principle of a PoCA algorithm. Actual (bold in red), and incoming and outgoing muon trajectories, $l_1$ and $l_2$, are illustrated. A PoCA voxel (blue) which is estimated from $l_1$ and $l_2$ is also shown in the voxelated volume of interest.

accordingly in a voxelated volume of interest, as shown in Fig. 1.

$$P_{PoCA}|_{3D} = (P_x, P_y, P_z) \rightarrow V_{N^3}(l,m,n). \qquad (1)$$

Once the PoCA voxel is determined, the scattering angle is assigned in that voxel to reconstruct images of a target object. Voxels are colored based on scattering angle values: rad or rad$^2$/cm. If the number of muon events is statistically significant, then colors become consistent with the material composition.

## MUON SCATTERING ANGLE DISTRIBUTION ESTIMATIONS

### Gaussian MCS Approximation

Muon scattering angle distribution provides the information needed to identify the property and size of a target object in MST. Muon scattering angle distribution is often approximated using a Gaussian distribution with a zero mean, and its width is determined using the Lynch and Dahl model [15],

$$\sigma_\theta = \frac{13.6 \text{ MeV}}{\beta c p} z \sqrt{\frac{X}{X_0}} \left[1 + 0.038 \ln\left(\frac{Xz^2}{X_0 \beta^2}\right)\right], \qquad (2)$$

where $\beta c$, $p$, and $z$ are the muon velocity, momentum, and charge number, $X$ is the length of scattering medium, and $X_0$ is the radiation length. $X_0$ for materials can be found in the library [4][16]. For unlisted materials, the radiation length can be approximated using a following formula [4]:

$$X_0[g/cm^3] = \frac{716.4 A}{Z(Z+1)\ln\left(\frac{287}{\sqrt{Z}}\right)}, \quad (3)$$

where $Z$ and $A$ are the charge and atomic number of materials, respectively.

**Molière Theory MCS Approximation**

A generalized form of the extended Molière model for MCS angle distribution within a scattering medium for a point-like nucleus can be written as [17]:

$$F(\theta, t)\theta d\theta = 2\chi e^{-\chi^2}\left[1 + \frac{b_0 + b_2\chi^2 + \sum_{\nu=2}^{\infty} b_{2\nu}\chi^{2\nu}}{B}\right]d\chi, \quad (4)$$

where $\theta$ is the muon scattering angle, and $t$ is the thickness of scattering medium. Definitions and details for each term can be found in [17][18]. Unlike Gaussian MCS approximation, the extended Molière model considers hard collision muon scattering events with a large angle in the second terms in the right-hand side of Eq. (4).

**GEANT4 Simulations**

In the Monte Carlo particle transport code GEANT4, a numerical model to simulate Coulombic interaction of charged particles in the scattering medium is developed based on the Goudsmit-Saunderson and Lewis models [19][20]. Muon scattering angle distribution is written using Legendre polynomials, $P_l(\cos\theta)$,

$$F(\theta, t) = \sum_{l=0}^{\infty} \frac{2l+1}{4\pi} e^{-\frac{X}{\lambda_l} P_l(\cos\theta)}, \quad (5)$$

where

$$\frac{1}{\lambda_l} = \exp\left(-2\pi X N \int_{-1}^{1}(1 - P_l(\cos\theta))\frac{d\sigma(\theta)}{d\Omega}d(\cos\theta)\right), \quad (6)$$

where $\lambda_l$ is the $i$th transport mean free path, $N$ is the atom number density of scattering medium, and $d\sigma(\theta)/d\Omega$ is a single scattering differential cross section initially derived by Rutherford [21]. A comparison of three MCS models–(1) the Gaussian MCS approximation, (2) the Molière model, and (3) the GEANT4 simulation for muon scattering angle distribution when $10^4$ 3 GeV muons interact with a 10 cm uranium sample–is shown in Fig 2. Unlike the MCS Gaussian approximation model, the Molière model successfully shows that its peak of distribution is located around 20 milliradian, and it has a long tail as a result of the large angle muon deflections.

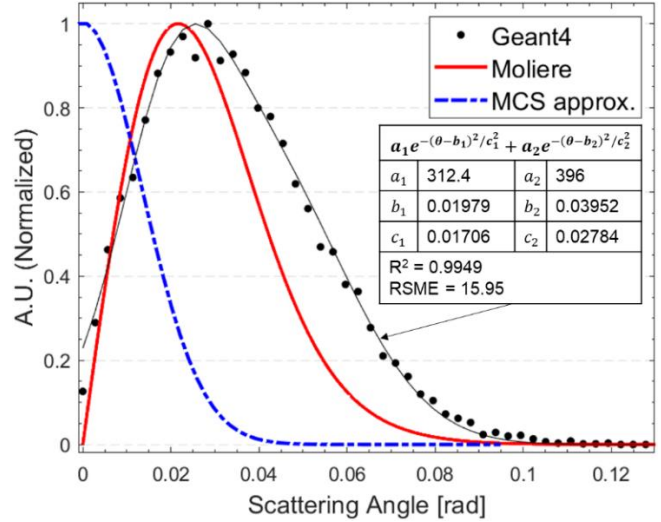

Fig. 2. Estimations of muon scattering angle distribution by three approaches: (1) Gaussian MCS approximation, (2) Molière model, and (3) GEANT4 simulation.

**MOMENTUM INTEGRATED POCA ALGORITHM**

As shown in Eq. (1), muon momentum affects the muon scattering angle distribution as significantly as material properties. To investigate the effect of muon momentum on distribution, $10^5$ poly-energetic muons were generated and allowed to interact with a $10\times10\times10$ cm$^3$ cubic tungsten sample in GEANT4. Then, the scattering angle data are decomposed by six momentum levels, 0.1–1.0, 1.0–2.0, …, and >5.0 GeV/c. The results in Fig. 3 shows that the most probable scattering angle, $mod(\theta)$, decreases as momentum increases.

To model a relationship between $mod(\theta)$ and muon momentum, simulations were repeated with $10^5$ various monoenergetic muons (1–100 GeV/c) and various sample materials (summarized in TABLE I).

TABLE I. Z number, size, and radiation length numbers for various sample materials

|    | Z  | Size [cm] | $X/X_0$  |
|----|----|-----------|----------|
| Al | 13 | 10        | 1.1240   |
| Fe | 26 | 10        | 5.6915   |
| Pb | 82 | 10        | 17.8190  |
| U  | 92 | 5         | 17.7928  |
| U  | 92 | 10        | 31.5856  |
| U  | 92 | 20        | 63.1712  |

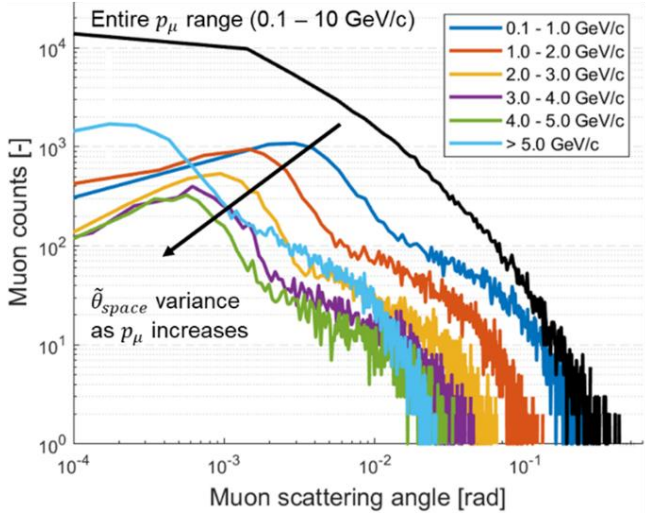

Fig. 3. Decomposed muon scattering angle distribution by six momentum levels when muons interact with 10×10×10 cm³ cubic tungsten sample.

## RESULTS

### M-value

The simulation results show that there is an exponential relationship between mode muon scattering angle, $mod(\theta)$, and muon momentum, $p$. Hence, its general relationship can be written by:

$$\log_{10} mod(\theta) = k\log_{10} p + M, \quad (7)$$

where $k$ is the slope and $M$ is the y-intercept. $k$ and $M$ were computed using GEANT4 simulations for various muon momenta and material samples. The results of Eq. (7) with $k$ and M-values for uranium, lead, iron, and aluminum samples are shown in Fig. 4.

In many applications, it is challenging to collect a statistically significant number of muon samples for each PoCA voxel, instead we can measure muon momentum and scattering for each muon event. Therefore, it is more practical to replace $mod(\theta)$ to $\theta$ in Eq. (7). Then, the M-value can be evaluated by:

$$M(p,\theta) \equiv \log_{10}\left(\frac{\theta}{p^k}\right). \quad (8)$$

Although the $k$-value varies depending on sizes and types of materials as shown in Fig. 4, it can be assumed as a constant because its variance is insignificant compared to that of M-values. For example, when $k$ is -2.3,

$$M = \log_{10}(\theta \ [rad] \times p \ [GeV/c]^{2.3}). \quad (9)$$

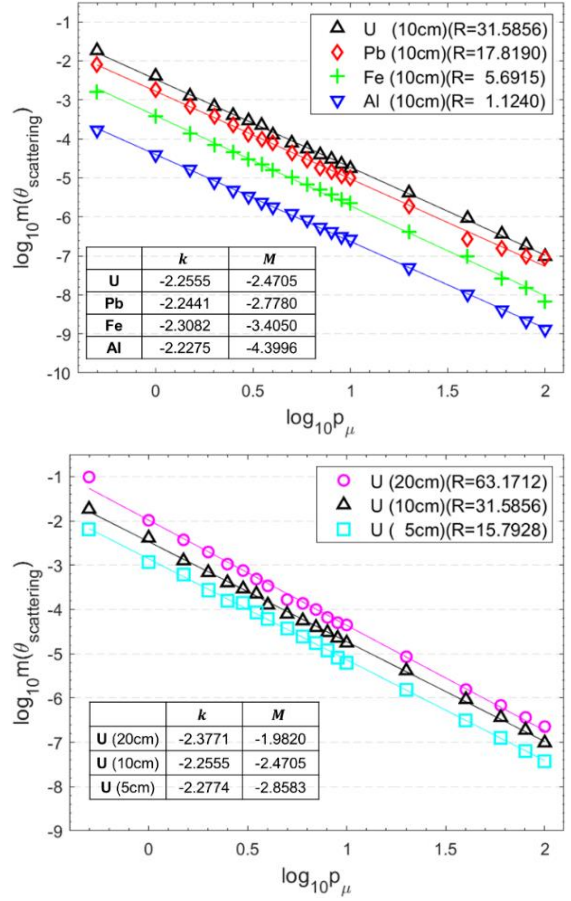

Fig. 4. Relationship between $mod(\theta)$ and muon momentum expressed by Eq. (7) for various sample materials, uranium, lead, iron, and aluminum (upper) and sizes of uranium, 5, 10, and 20 cm (lower). All samples have unique radiation length numbers, $R$. $k$ (slope) and M-values (y-intercept) for all materials are summarized in figures.

### Application to Spent Nuclear Fuel Cask Imaging

MST can be used to monitor SNF casks. A typical SNF assembly in dry storage was simulated in GEANT4 to investigate the benefit of using M-value in MST. In a current model, one of 24 FAs was missing in storage, as shown in Fig. 5a. Tomographic images of the SNF assemblies were reconstructed by applying two MST algorithms: (1) PoCA with scattering angle (Fig. 5b) and (2) momentum-dependent PoCA with M-values (Figs. 5c and 5d). The results are significantly improved by replacing the scattering angles with M-values, enabling the location of a missing FA that was not visible without measuring muon momentum. Fine muon momentum measurement enhances tomography quality. Figs. 5c and 5d show the images that were reconstructed using M-values with momentum uncertainties, $\sigma_p$, of 0.1 and 0.01 GeV/c, respectively.

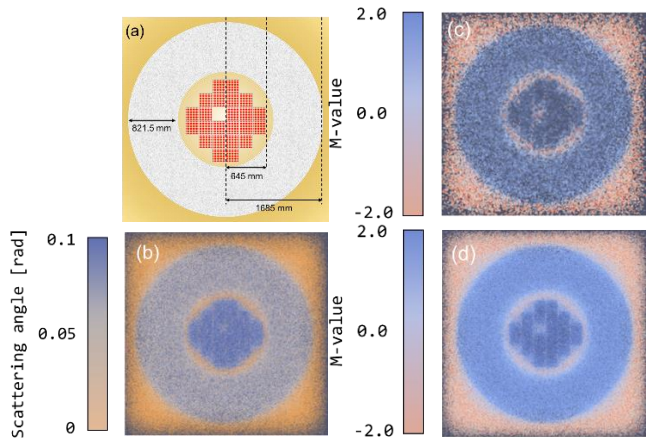

Fig. 5. Placements of SNF assemblies in a dry storage when one FA is missing in the center, (a) a model and reconstructed cross-section images using (b) muon scattering angle, and M-values with (c) $\sigma_p$=0.1 GeV/c and (d) $\sigma_p$=0.01 GeV/c uncertainties.

## CONCLUSION

The benefit of measuring muon momentum in MST is presented in this paper. The relationships between muon momentum and scattering angle distribution in various material samples are thoroughly investigated using Gaussian MSC approximation, Molière model approximation, and GEANT4 simulation. Although Molière model successfully predicts the hard collision muon scattering, it underestimates the frequency of muon events with a large angle. GEANT4 simulations have been utilized to develop a new imaging algorithm integrating muon scattering angle and momentum in a single value, or M-value. A tomographic image of SNF assemblies in dry storage was reconstructed using PoCA with M-value algorithms. The results show that the image quality is significantly improved when muon momentum information is coupled with scattering angle value, and the image is much enhanced by using a muon momentum measurement.

## ACKNOWLEDGMENTS

This research was sponsored by the Laboratory Directed Research and Development Program of Oak Ridge National Laboratory, managed by UT-Battelle, LLC, for the US Department of Energy.